\documentclass[12pt]{article}

\usepackage{epsfig,amsmath,amsthm,amssymb,amsfonts,xypic,graphicx}
\begin{document}

\newtheorem{thm}{Theorem}[section]
\newtheorem{lem}[thm]{Lemma}
\newtheorem{cor}[thm]{Corollary}
\newtheorem{prop}[thm]{Proposition}
\newtheorem{remark}[thm]{Remark}
\newtheorem{defn}[thm]{Definition}
\newtheorem{ex}[thm]{Example}
\newtheorem{conj}[thm]{Conjecture}
\newtheorem{prop-conj}[thm]{Proposition-Conjecture}
\newtheorem{prop-defn3}[thm]{Proposition-Definition3}
\newtheorem{prop-defn}[thm]{Proposition-Definition}
\newtheorem{defn1}[thm]{Supplementary Definition}
\newtheorem{defn2}[thm]{Definition2}
\newenvironment{ack}{Acknowledgements}

\numberwithin{equation}{section}
\newcommand{\mc}{\mathcal}
\newcommand{\mb}{\mathbb}
\newcommand{\surj}{\twoheadrightarrow}
\newcommand{\inj}{\hookrightarrow}
\newcommand{\red}{{\rm red}}
\newcommand{\codim}{{\rm codim}}
\newcommand{\rank}{{\rm rank}}
\newcommand{\Pic}{{\rm Pic}}
\newcommand{\Div}{{\rm Div}}
\newcommand{\Hom}{{\rm Hom}}
\newcommand{\im}{{\rm im}}
\newcommand{\Sym}{{\rm Sym}}
\newcommand{\Ker}{{\rm Ker}}
\newcommand{\Spec}{{\rm Spec \,}}
\newcommand{\Sing}{{\rm Sing}}
\newcommand{\Char}{{\rm char}}
\newcommand{\Tr}{{\rm Tr}}
\newcommand{\Gal}{{\rm Gal}}
\newcommand{\Min}{{\rm Min \ }}
\newcommand{\Max}{{\rm Max \ }}
\newcommand{\Ext}{{\rm Ext }}
\newcommand{\End}{{\rm End }}
\newcommand{\Tor}{{\rm Tor }}
\newcommand{\sgn}{\text{sgn}}
\newcommand{\sA}{{\mathcal A}}
\newcommand{\sB}{{\mathcal B}}
\newcommand{\sC}{{\mathcal C}}
\newcommand{\sD}{{\mathcal D}}
\newcommand{\sE}{{\mathcal E}}
\newcommand{\sF}{{\mathcal F}}
\newcommand{\sG}{{\mathcal G}}
\newcommand{\sH}{{\mathcal H}}
\newcommand{\sI}{{\mathcal I}}
\newcommand{\sJ}{{\mathcal J}}
\newcommand{\sK}{{\mathcal K}}
\newcommand{\sL}{{\mathcal L}}
\newcommand{\sM}{{\mathcal M}}
\newcommand{\sN}{{\mathcal N}}
\newcommand{\sO}{{\mathcal O}}
\newcommand{\sP}{{\mathcal P}}
\newcommand{\sQ}{{\mathcal Q}}
\newcommand{\sR}{{\mathcal R}}
\newcommand{\sS}{{\mathcal S}}
\newcommand{\sT}{{\mathcal T}}
\newcommand{\sU}{{\mathcal U}}
\newcommand{\sV}{{\mathcal V}}
\newcommand{\sW}{{\mathcal W}}
\newcommand{\sX}{{\mathcal X}}
\newcommand{\sY}{{\mathcal Y}}
\newcommand{\sZ}{{\mathcal Z}}

\newcommand{\fM}{{\frak M}}
\newcommand{\fD}{{\frak D}}
\newcommand{\fH}{{\frak H}}
\newcommand{\fG}{{\frak G}}
\newcommand{\fU}{{\frak U}}
\newcommand{\fg}{{\frak g}}
\newcommand{\fu}{{\frak u}}

\newcommand{\A}{{\Bbb A}}
\newcommand{\B}{{\Bbb B}}
\newcommand{\C}{{\Bbb C}}
\newcommand{\D}{{\Bbb D}}
\newcommand{\E}{{\Bbb E}}
\newcommand{\F}{{\Bbb F}}
\newcommand{\G}{{\Bbb G}}
\renewcommand{\H}{{\Bbb H}}
\newcommand{\I}{{\Bbb I}}
\newcommand{\J}{{\Bbb J}}
\newcommand{\M}{{\Bbb M}}
\newcommand{\N}{{\Bbb N}}
\renewcommand{\P}{{\Bbb P}}
\newcommand{\Q}{{\Bbb Q}}
\newcommand{\R}{{\Bbb R}}
\newcommand{\T}{{\Bbb T}}
\newcommand{\U}{{\Bbb U}}
\newcommand{\V}{{\Bbb V}}
\newcommand{\W}{{\Bbb W}}
\newcommand{\X}{{\Bbb X}}
\newcommand{\Y}{{\Bbb Y}}
\newcommand{\Z}{{\Bbb Z}}
\def\mycases#1#2{\left.\vcenter{\hbox{$#1$}\hbox{$#2$}}\right.}
\title{Energy-Momentum of a regular MMaS-class black hole}  
\author{Owen Patashnick\footnote{Partially supported by the Center for Advanced Studies in Mathematics, Ben Gurion University, Be'er Sheva, Israel, and by the National Planning and Grant Committee (Israel)} \quad powen@math.bgu.ac.il} 


\maketitle

\begin{abstract}

We compute the energy and momentum of a 
regular black hole of type defined by Mars, Mart\'{\i}n-Prats, and Senovilla
using the Einstein and Papapetrou definitions for energy-momentum density.  
Some other definitions of energy-momentum density are shown to give mutually contradictory and less reasonable results.
Results support the Cooperstock hypothesis.
\end{abstract}



\section{Introduction}\label{intro:begin}

An important outstanding issue in the General Theory of Relativity is the problem of finding an acceptable definition of energy-momentum localization.

H. Bondi \cite{Bon} has argued that General Relativity does not permit a non-localizible form of energy, so in principle, an acceptable definition of local energy-momentum density should exist.

The first attempts to identify such a density by Einstein himself \cite{E} as well as Landau and Lifshitz \cite{LL}, Papapetrou \cite{P}, and Weinberg \cite{W}, were all defined in terms of non-tensorial, coordinate-dependent pseudotensors. 
Unfortunately, the physical meaning of the pseudotensors has been and remains a serious point of contention.  (For example, one train of thought suggests that a pseudotensor approach could conflict with the equivalence principle (\cite{MKW})).

Attempts to find alternative definitions of a local energy-momentum 
density
(by M\o ller \cite{Mo}, Komar \cite{Ko}, Penrose \cite{Pen}, and others), however, 
have not been substantially more successful;
for example, Bergqvist \cite{Ber} showed that no two of the many coordinate-independent definitions of mass give the same result for the Reissner-Nordstrom and Kerr spacetimes.

By contrast, 
Aguirregabiria, Chamorro, and Virbhadra \cite{ACV} showed that for 
all spacetimes of Kerr-Schild class,
the energy-momentum complexes 
defined by 
Einstein, Landau-Lifshitz, Papapetrou and Weinberg all give the same 
answer.  In \cite{V1} Virbhadra noticed that the abovementioned energy-momentum complexes coincide for a class of solutions more general than Kerr-Schild.
For example, Rosen and Virbhadra \cite{RV} showed that several energy-momentum complexes give the same and a meaningful result for the Einstein-Rosen metric describing cylindrical gravitational waves; note that this metric is not of Kerr-Schild class.

Furthermore, Chang, Nester and Chen \cite{CNC1}, \cite{CNC2} showed that these 
\linebreak
energy-momentum complexes can be redefined as the value of a Hamiltonian, which has a surface and a boundary term.  The pseudotensors of Einstein, Landau-Lifshitz, Papapetrou, Weinberg and M\o ller can then be defined in terms of the boundary term of the Hamiltonian.  Thus the pseudotensors are in fact quasi-local, and under this formulation the energy-momentum complexes no longer conflict with the equivalence principle.  

As a result there has been a renaissance of interest in and use of one or more energy-momentum complexes in a wide variety of contexts \cite{list}.

In general the various energy-momentum complexes give different answers 
\cite{V1}, \cite{Xulu}.
Chang, Nestor and Chen \cite{CNC1} pointed out that the physical significance of the boundary term is still unclear, and hence there are {\it a priori} many choices possible for an appropriate boundary term.

It is therefore natural to ask if any of the various definitions of quasi-local energy-momentum density is preferable to the others.







In this paper we calculate the energy and momentum distributions of
a class of regular black holes defined by Mars, Mart\'{\i}n-Pratt, and Senovilla \cite{MMS} using several of the complexes mentioned above.
We will show that the Einstein energy-momentum complex is one of the complexes that gives a meaningful answer for the energy and momentum distributions, supporting the hypothesis (\cite{V1},\cite{RV}) that the Einstein energy-momentum complex is a reasonable definition of energy-momentum density.






We use the convention that Latin indices take values from 0 to 3 and Greek indices take values from 1 to 3, and we take units where $G=1$ and $c=1$.  The comma and semicolon, respectively, stand for partial and covariant derivatives.

\section{The Mars--Mart\'{\i}n-Pratt--Senovilla regular black hole}

Although the singularity theorem of Hawking and Ellis \cite{HE} forbids the existence of regular black holes with matter content satisfying the strong energy condition everywhere, it is still possible to construct regular black holes satisfying the so called weak energy condition everywhere (the energy-momentum tensor $T_{a b}$ must satisfy $T_{a b}W^{a}W^{b}\geq 0$ where $W$ is any timelike vector at any point of spacetime.)
A famous example of such a regular black hole was given by Bardeen \cite{Bar}.  (Also see Sharifi \cite{Shar} for the energy computation).  In this paper we consider another class of regular black holes constructed by Mars, Mart\'{\i}n-Pratt and Senovilla \cite{MMS}.
 
Consider the following line element:

\begin{eqnarray}\label{MMaS} 
ds^2&=&e^{4\beta(r)}\chi(r)du^2-2e^{2\beta(r)}dudr-r^2(d\theta^2+\sin^2(\theta)d\phi^2)\end{eqnarray}
where
\begin{eqnarray}
\chi(r)&=&1-\frac{2M(r)}{r}
\end{eqnarray}

The functions $M$ and $\beta$ of $r$ are chosen so as to satisfy the 
following three constraints:
\begin{enumerate}
\item the spacetime must represent a Schwarzchild black hole,
(i.e. $$ M(r)=m\quad\quad \beta'(r)=0\quad\quad \text {for }r>r_0 \text{ with } r_0\leq 2m$$ where m is some positive constant which represents the total mass of the black hole, and prime denotes the derivative with respect to $r$), 
\item the model fulfills the weak energy conditions everywhere, and
\item the spacetime is regular.
\end{enumerate}

In the paper \cite {MMS} the authors give two explicit examples of models that fulfill all these conditions (hence models do exist).  In this paper we choose to leave $M$ and $\beta$ general in all of our computations; the computations are all done for an arbitrary choice of functions $M$ and $\beta$ of $r$ that satisfy the above criteria.  However, for the graphs we will use the following solution from \cite {MMS}:
\begin{eqnarray}
M(x)=mx^3(10-15x+6x^2)
\end{eqnarray}
where $x, 0\leq x\leq 1$ is the adimensional variable $x=r/2m$, and the 
corresponding function $\beta$ can be integrated from
\begin{eqnarray}
\beta'(x)=5(1-x)^2x
\end{eqnarray}

Note that the line element \eqref{MMaS} is a modification of the \linebreak 
Eddington-Finkelstein form of a Schwarzschild-like spacetime 
\begin{eqnarray}\label{Schwartz}
ds^2=-\chi(r)du^2+2dudr+r^2(d\theta^2+\sin^2(\theta)d\phi^2)
\end{eqnarray}
by $e^{2\beta(r)}$ in the ``$u$'' direction, since the (unorthodox) transformation \linebreak $u \rightarrow e^{-2\beta(r)}u$ gets rid of this term.  Hence, the energy of the interior of a black hole defined by the line element \eqref{MMaS} should be modified from the energy of a black hole defined by \eqref{Schwartz} by the scalar field $e^{2\beta(r)}$.

Thus from physical considerations we expect the energy of the region enclosed in a 2-sphere to be given by the product $M(r)e^{2\beta(r)}$ of the mass function $M(r)$ and the scalar field $e^{2\beta(r)}$.

In order to calculate the energy and momentum distributions in the next two sections we will need to rewrite \eqref{MMaS} in cartesian coordinates.

Transforming the line element \eqref{MMaS} to cartesian coordinates via
\begin{eqnarray}
u&=&t+r\\
x&=&r\sin(\theta)\cos(\phi)\\
y&=&r\sin(\theta)\sin(\phi)\\
z&=&r\cos(\theta)
\end{eqnarray}
yields
\begin{multline}\label{MMSlineelt}
ds^2=-dx^2-dy^2-dz^2+\frac{(1-e^{2\beta(r)})}{r^2}[xdx+ydy+zdz]^2\\+e^{2\beta(r)}\Big[dt^2-(1-e^{2\beta(r)}\chi(r))[dt+\frac{1}{r}(xdx+ydy+zdz)]^2\Big]
\end{multline}

Note that \eqref{MMaS} (equivalently \eqref{MMSlineelt} ) is a static spherically symmetric metric, but is {\bf not} a Kerr-Schild class metric.  In fact, \eqref{MMaS} falls outside the class of metrics for which the energy-momentum complexes of Einstein, Landau-Lifshitz, Papapetrou, and Weinberg coincide.  We will demonstrate this explicitly in the following two sections.

\section{Energy-Momentum using the Einstein Energy-Momentum complex}

The Einstein energy-momentum complex is 
\begin{eqnarray}\label{Edensity}
\Theta_i^k&=&\frac{1}{16\pi}H_i^{kl}{}_{,l}
\end{eqnarray}
where 
\begin{eqnarray}\label{Edensity'}
H_i^{kl}=-H_i^{lk}=\frac{g_{in}}{\sqrt{-g}}[-g(g^{kn}g^{lm}-g^{ln}g^{km})]_{,m}
\end{eqnarray}
$\Theta_0^0$ and $\Theta_{\alpha}^0$ denote energy and momentum density respectively.  The energy and momentum components are given by
\begin{eqnarray}\label{Ecomps}
P_i=\int\int\int\Theta_i^0dx^1dx^2dx^3=\frac{1}{16\pi}\int\int H^{0\alpha}_{i}n_{\alpha}dS
\end{eqnarray}
where we applied Gauss's theorem to get the second equality ($P_0$ and $P_{\alpha}$ stand for the energy and momentum components respectively, and $n_{\alpha}$ denotes the outward unit normal to the infinitesimal surface element $dS$).

Using \eqref{MMSlineelt} in \eqref{Edensity'}, it is easy to compute  
the twelve superpotentials $H_i^{\alpha,0}$ as follows:
\begin{eqnarray}
H_0^{01}=\frac{-2x}{r^2}e^{2\beta(r)}(-1+\chi(r))\\
H_0^{02}=\frac{-2y}{r^2}e^{2\beta(r)}(-1+\chi(r))\\
H_0^{03}=\frac{-2z}{r^2}e^{2\beta(r)}(-1+\chi(r))
\end{eqnarray}
\begin{multline}
H_1^{01}=\frac{1}{r^3}\big{(}x^2(2-2e^{2\beta(r)}\chi(r))\\-(y^2+z^2)(1-2r\beta'(r)+e^{2\beta(r)}\chi(r)(1+4r\beta'(r))+re^{2\beta(r)}\chi'(r))\big{)}
\end{multline}
\begin{multline}
H_1^{02}=\frac{1}{r^3}\big{(}xy(1-2r\beta'(r)+e^{2\beta(r)}\chi(r)(-1+4r\beta'(r))+e^{2\beta(r)}r\chi'(r))\big{)}
\end{multline}
\begin{multline}
H_1^{03}=\frac{1}{r^3}\big{(}xz(1-2r\beta'(r)+e^{2\beta(r)}\chi(r)(-1+4r\beta'(r))+e^{2\beta(r)}r\chi'(r))\big{)}
\end{multline}
\begin{multline}
H_2^{01}=\frac{1}{r^3}\big{(}xy(1-2r\beta'(r)+e^{2\beta(r)}\chi(r)(-1+4r\beta'(r))+e^{2\beta(r)}r\chi'(r))\big{)}
\end{multline}
\begin{multline}
H_2^{02}=\frac{1}{r^3}\big{(}y^2(2-2e^{2\beta(r)}\chi(r))\\-(x^2+z^2)(-1-2r\beta'(r)+e^{2\beta(r)}\chi(r)(1+4r\beta'(r))+re^{2\beta(r)}\chi'(r))\big{)}
\end{multline}
\begin{multline}
H_2^{03}=\frac{1}{r^3}\big{(}yz(1-2r\beta'(r)+e^{2\beta(r)}\chi(r)(-1+4r\beta'(r))+e^{2\beta(r)}r\chi'(r))\big{)}
\end{multline}
\begin{multline}
H_3^{01}=\frac{1}{r^3}\big{(}xz(1-2r\beta'(r)+e^{2\beta(r)}\chi(r)(-1+4r\beta'(r))+e^{2\beta(r)}r\chi'(r))\big{)}
\end{multline}
\begin{multline}
H_3^{02}=\frac{1}{r^3}\big{(}yz(1-2r\beta'(r)+e^{2\beta(r)}\chi(r)(-1+4r\beta'(r))+e^{2\beta(r)}r\chi'(r))\big{)}
\end{multline}
\begin{multline}
H_3^{03}=\frac{1}{r^3}\big{(}z^2(2-2e^{2\beta(r)}\chi(r))\\-(x^2+y^2)(1-2r\beta'(r)+e^{2\beta(r)}\chi(r)(1+4r\beta'(r))+re^{2\beta(r)}\chi'(r))\big{)}
\end{multline}

(here prime denotes derivative with respect to $r$).

Now applying \eqref{Ecomps} one gets 
\begin{eqnarray}\label{Eenergy} E=e^{2\beta(r)}M(r) \end{eqnarray} 
for the energy component.  Note that \eqref{Eenergy} is the value for $E$ we deduced in section 2.
Furthermore note that for $r\geq2m$, \eqref{Eenergy} reduces to $E=m$, the energy expression for a Schwarzchild black hole.

Using \eqref{Eenergy} and recalling our specific choice of model in section 2, in Figure 1 we plot $E/m$ (on the $y$-axis) versus $r/2m$. 

Similarly, the momentum components are:

\begin{eqnarray}\label{Emomenta} 
P_1=P_2=P_3=0
\end{eqnarray}

as expected for a static spherically symmetric black hole.

\begin{figure}[htbp]
\begin{center}
\includegraphics[scale=1]{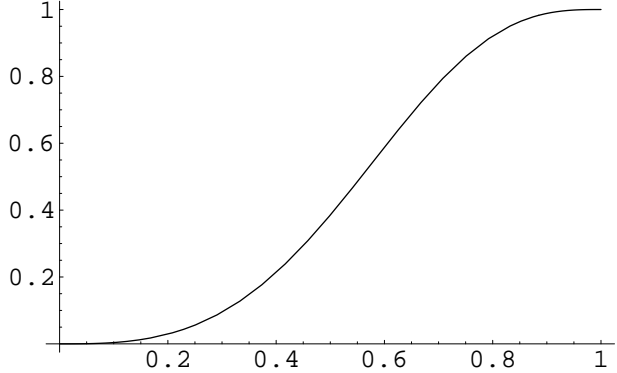}
\end{center}
\vspace{-.3in}
\caption{E/m vs. r/2m for \eqref{Eenergy}}
\label{fig1}
\end{figure}

\section{Energy computations involving other Energy-Momentum complexes}

The Landau-Lifshitz energy-momentum complex is 
\begin{eqnarray}\label{LLdensity}
L^{ik}&=&\frac{1}{16\pi}\lambda^{iklm}{}_{,lm}
\end{eqnarray}
where 
\begin{eqnarray}\label{LLdensity'}
\lambda^{iklm}&=&-g(g^{ik}g^{lm}-g^{il}g^{km})
\end{eqnarray}
$L^{00}$ and $L^{\alpha 0}$ denote energy and momentum density respectively.  The energy and momentum components are given by
\begin{eqnarray}\label{LLcomps}
P_i=\int\int\int L^{i0}dx^1dx^2dx^3=\frac{1}{16\pi}\int\int \lambda^{i0\alpha m}{}_{,m}n_{\alpha}dS
\end{eqnarray}
where we applied Gauss's theorem to get the second equality ($P_0$ and $P_{\alpha}$ stand for the energy and momentum components respectively, and $n_{\alpha}$ denotes the outward unit normal to the infinitesimal surface element $dS$).

Using \eqref{MMSlineelt} in \eqref{LLdensity'}, it is easy to compute  
the twelve superpotentials $\lambda^{i0\alpha}$ as follows:
\begin{eqnarray}
\lambda^{001}&=&\frac{2x}{r^2}(1-2e^{2\beta(r)}+e^{4\beta(r)}\chi(r))\\
\lambda^{002}&=&\frac{2y}{r^2}(1-2e^{2\beta(r)}+e^{4\beta(r)}\chi(r))\\
\lambda^{003}&=&\frac{2z}{r^2}(1-2e^{2\beta(r)}+e^{4\beta(r)}\chi(r))
\end{eqnarray}
\begin{multline}
\lambda^{101}=  \frac{1}{r^3}\big{(}e^{2\beta(r)}(2x^2(-1+e^{2\beta(r)}\chi(r))\\+(y^2+z^2)(-1-2r\beta'(r)+e^{2\beta(r)}\chi(r)(1+4r\beta'(r))+re^{2\beta(r)}\chi'(r)))\big{)}
\end{multline}
\begin{multline}
\lambda^{102}=-\frac{1}{r^3}\big{(}xye^{2\beta(r)}(1-2r\beta'(r)+e^{2\beta(r)}\chi(r)(-1+4r\beta'(r))+e^{2\beta(r)}r\chi'(r))\big{)}
\end{multline}
\begin{multline}
\lambda^{103}=-\frac{1}{r^3}\big{(}xze^{2\beta(r)}(1-2r\beta'(r)+e^{2\beta(r)}\chi(r)(-1+4r\beta'(r))+e^{2\beta(r)}r\chi'(r))\big{)}
\end{multline}
\begin{multline}
\lambda^{201}=-\frac{1}{r^3}\big{(}xye^{2\beta(r)}(1-2r\beta'(r)+e^{2\beta(r)}\chi(r)(-1+4r\beta'(r))+e^{2\beta(r)}r\chi'(r))\big{)}
\end{multline}
\begin{multline}
\lambda^{202}= \frac{1}{r^3}\big{(}e^{2\beta(r)}(2y^2(-1+e^{2\beta(r)}\chi(r))\\+(x^2+z^2)(-1-2r\beta'(r)+e^{2\beta(r)}\chi(r)(1+4r\beta'(r))+re^{2\beta(r)}\chi'(r)))\big{)}
\end{multline}
\begin{multline}
\lambda^{203}=-\frac{1}{r^3}\big{(}yze^{2\beta(r)}(1-2r\beta'(r)+e^{2\beta(r)}\chi(r)(-1+4r\beta'(r))+e^{2\beta(r)}r\chi'(r))\big{)}
\end{multline}
\begin{multline}
\lambda^{301}=-\frac{1}{r^3}\big{(}xze^{2\beta(r)}(1-2r\beta'(r)+e^{2\beta(r)}\chi(r)(-1+4r\beta'(r))+e^{2\beta(r)}r\chi'(r))\big{)}
\end{multline}
\begin{multline}
\lambda^{302}=-\frac{1}{r^3}\big{(}yze^{2\beta(r)}(1-2r\beta'(r)+e^{2\beta(r)}\chi(r)(-1+4r\beta'(r))+e^{2\beta(r)}r\chi'(r))\big{)}
\end{multline}
\begin{multline}
\lambda^{303}= \frac{1}{r^3}\big{(}e^{2\beta(r)}(2z^2(-1+e^{2\beta(r)}\chi(r))\\+(x^2+y^2)(-1-2r\beta'(r)+e^{2\beta(r)}\chi(r)(1+4r\beta'(r))+re^{2\beta(r)}\chi'(r)))\big{)}
\end{multline}

Now applying \eqref{LLcomps} one gets 
\begin{eqnarray}\label{LLenergy} E=-\frac{r}{2}(1-2e^{2\beta(r)}+e^{4\beta(r)}(1-\frac{2M(r)}{r})) \end{eqnarray} 
for the energy component, and 
\begin{eqnarray}\label{LLmomenta} 
P_1=P_2=P_3=0
\end{eqnarray}
for the momentum components
.  Although formula \eqref{LLenergy} reduces to the correct expression ($E=m$) for the region $r\geq 2m$, the energy component \eqref{LLenergy} differs from \eqref{Eenergy} for $r<2m$.  Thus \eqref{MMaS} falls outside the class of metrics for which the energy-momentum complexes of Einstein, Landau-Lifshitz, Papapetrou, and Weinberg coincide.  Furthermore, 
negative energy is not physically meaningful for electrically neutral nonrotating massive objects, while the energy component \eqref{LLenergy} is negative for small $r$. (see Figure 2, where we plot $E/m$ (on the $y$-axis) versus $r/2m$.)  Therefore, based on physical considerations, \eqref{LLenergy} appears to be incorrect for $r<2m$, which supports the contention that the Einstein energy-momentum complex is a more reasonable computational tool.

\begin{figure}[htbp]
\begin{center}
\includegraphics[scale=1]{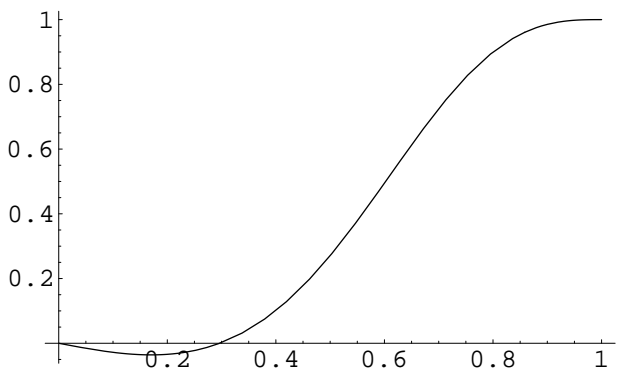}
\end{center}
\vspace{-.3in}
\caption{E/m vs. r/2m for \eqref{LLenergy}}
\label{fig2}
\end{figure}

The M\o ller energy-momentum complex is 
\begin{eqnarray}\label{Mdensity}
J_i^k&=&\frac{1}{8\pi}\sJ_{i,l}^{kl}
\end{eqnarray}
where 
\begin{eqnarray}\label{Mdensity'}
\sJ_{i}^{kl}=\sqrt{-g}(g_{ia,b}-g_{ib,a})g^{kb}g^{la}
\end{eqnarray}
$J_0^0$ and $J_{\alpha}^0$ denote energy and momentum density respectively.  The energy and momentum components are given by
\begin{eqnarray}\label{Mcomps}
P_i=\int\int\int J_i^0dx^1dx^2dx^3=\frac{1}{8\pi}\int\int \sJ^{0\alpha}_{i}n_{\alpha}dS
\end{eqnarray}
where we applied Gauss's theorem to get the second equality ($P_0$ and $P_{\alpha}$ stand for the energy and momentum components respectively, and $n_{\alpha}$ denotes the outward unit normal to the infinitesimal surface element $dS$).

Using \eqref{MMSlineelt} in \eqref{Mdensity'}, it is easy to compute  
the twelve superpotentials $\sJ_i^{0\alpha}$ as follows:
\begin{eqnarray}
\sJ_0^{01}&=&\frac{xe^{2\beta(r)}}{r^2}\big{(}4\chi(r)\beta'(r)+\chi'(r)\big{)}\\
\sJ_0^{02}&=&\frac{ye^{2\beta(r)}}{r^2}\big{(}4\chi(r)\beta'(r)+\chi'(r)\big{)}\\
\sJ_0^{03}&=&\frac{ze^{2\beta(r)}}{r^2}\big{(}4\chi(r)\beta'(r)+\chi'(r)\big{)}
\end{eqnarray}
\begin{multline}
\sJ_1^{01}=\frac{1}{r^3}\big{(}rx^2((-2+4e^{2\beta(r)}\chi(r))\beta'(r)+e^{2\beta(r)}\chi'(r))+(y^2+z^2)(-1+e^{2\beta(r)}\chi(r))\big{)}
\end{multline}
\begin{multline}
\sJ_1^{02}=\frac{xy}{r^3}\big{(}1-2r\beta'(r)+e^{2\beta(r)}\chi(r)(-1+4r\beta'(r))+re^{2\beta(r)}\chi'(r)\big{)}
\end{multline}
\begin{multline}
\sJ_1^{03}=\frac{xz}{r^3}\big{(}1-2r\beta'(r)+e^{2\beta(r)}\chi(r)(-1+4r\beta'(r))+re^{2\beta(r)}\chi'(r)\big{)}
\end{multline}
\begin{multline}
\sJ_2^{01}=\frac{xy}{r^3}\big{(}1-2r\beta'(r)+e^{2\beta(r)}\chi(r)(-1+4r\beta'(r))+re^{2\beta(r)}\chi'(r)\big{)}
\end{multline}
\begin{multline}
\sJ_2^{02}=\frac{1}{r^3}\big{(}ry^2((-2+4e^{2\beta(r)}\chi(r))\beta'(r)+e^{2\beta(r)}\chi'(r))+(x^2+z^2)(-1+e^{2\beta(r)}\chi(r))\big{)}
\end{multline}
\begin{multline}
\sJ_2^{03}=\frac{yz}{r^3}\big{(}1-2r\beta'(r)+e^{2\beta(r)}\chi(r)(-1+4r\beta'(r))+re^{2\beta(r)}\chi'(r)\big{)}
\end{multline}
\begin{multline}
\sJ_3^{01}=\frac{xz}{r^3}\big{(}1-2r\beta'(r)+e^{2\beta(r)}\chi(r)(-1+4r\beta'(r))+re^{2\beta(r)}\chi'(r)\big{)}
\end{multline}
\begin{multline}
\sJ_3^{02}=\frac{yz}{r^3}\big{(}1-2r\beta'(r)+e^{2\beta(r)}\chi(r)(-1+4r\beta'(r))+re^{2\beta(r)}\chi'(r)\big{)}
\end{multline}
\begin{multline}
\sJ_3^{03}=\frac{1}{r^3}\big{(}rz^2((-2+4e^{2\beta(r)}\chi(r))\beta'(r)+e^{2\beta(r)}\chi'(r))+(x^2+y^2)(-1+e^{2\beta(r)}\chi(r))\big{)}
\end{multline}

Now applying \eqref{Mcomps} one gets 
\begin{eqnarray}\label{Menergy} E=\frac{r^2}{2}e^{2\beta(r)}(4(1-\frac{2M(r)}{r})\beta'(r)-2(\frac{M'(r)r-M(r)}{r^2}))
\end{eqnarray} 
for the energy component (Here prime denotes derivative with respect to $r$), and 
\begin{eqnarray}\label{Mmomenta} 
P_1=P_2=P_3=0
\end{eqnarray}
for the momentum components.
Although again formula \eqref{Menergy} reduces to the correct expression ($E=m$) for the region $r\geq 2m$, the energy component \eqref{Menergy} differs from \eqref{Eenergy} for $r<2m$ 
.
(See Figure 3, where
we plot $E/m$ (on the $y$-axis) versus $r/2m$.)
Therefore, based on comparison with our expected answer, \eqref{Menergy} is less reasonable for $r<2m$, which supports the contention that the Einstein energy-momentum complex is a more reasonable computational tool.

\begin{figure}[htbp]
\begin{center}
\includegraphics[scale=1]{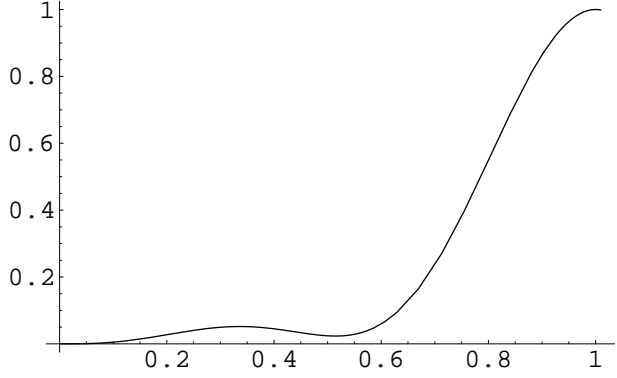}
\end{center}
\vspace{-.3in}
\caption{E/m vs. r/2m for \eqref{Menergy}}
\label{fig3}
\end{figure}

The Papapetrou energy-momentum complex is 
\begin{eqnarray}\label{Pdensity}
\Sigma^{ik}&=&\frac{1}{16\pi}N^{iklm}{}_{,lm}
\end{eqnarray}
where 
\begin{eqnarray}\label{Pdensity'}
N^{iklm}=\sqrt{-g}(g^{ik}\eta^{lm}-g^{il}\eta^{km}+g^{lm}\eta^{ik}-g^{lk}\eta^{im})
\end{eqnarray}
and $$\eta^{ik}=diag(1,-1,-1,-1)$$ 
$\Sigma^{00}$ and $\Sigma^{\alpha 0}$ denote energy and momentum density respectively.  The energy and momentum components are given by
\begin{eqnarray}\label{Pcomps}
P_i=\int\int\int\Sigma^{i0} dx^1dx^2dx^3=\frac{1}{16\pi}\int\int N^{i0\alpha\beta}_{,\beta} n_{\alpha} dS
\end{eqnarray}
where we applied Gauss's theorem to get the second equality ($P_0$ and $P_{\alpha}$ stand for the energy and momentum components respectively, and $n_{\alpha}$ denotes the outward unit normal to the infinitesimal surface element $dS$).  Note that we can apply Gauss' theorem since the metric \eqref{MMaS} (equivalently \eqref{MMSlineelt}) is time-independent.

Using \eqref{MMSlineelt} in \eqref{Pdensity'}, it is easy to compute  
the only three nonzero superpotentials $N^{i0\alpha}$ as follows:
\begin{eqnarray}
N^{001}=\frac{-2x}{r^2}e^{2\beta(r)}(-1+\chi(r))\\
N^{002}=\frac{-2y}{r^2}e^{2\beta(r)}(-1+\chi(r))\\
N^{003}=\frac{-2z}{r^2}e^{2\beta(r)}(-1+\chi(r))
\end{eqnarray}
since 
\begin{eqnarray}
N^{i0\alpha}=0 \text{ for } i>0.
\end{eqnarray}

Now applying \eqref{Pcomps} one gets the correct expression 
\begin{eqnarray}\label{Penergy} E=e^{2\beta(r)}M(r) \end{eqnarray} 
for the energy component, and   
\begin{eqnarray}\label{Pmomenta} 
P_1=P_2=P_3=0
\end{eqnarray}
for the momentum components, which agree with the expected values.  Thus the Papapetrou energy-momentum complex also appears to be a reasonable definition for energy-momentum density.

\section{The Cooperstock hypothesis}
The results of this paper support the Cooperstock hypothesis.
The Cooperstock hypothesis \cite{Coo} states that energy is localized to the region where the energy-momentum tensor is non-vanishing.  This hypothesis would imply that there is no energy-momentum contribution from ``vacuum'' regions of space-time.  If true, this hypothesis would have broad implications.  For example, the hypothesis suggests that gravitational waves have no energy and that current attempts to detect these waves using bar detectors are doomed to failure\footnote{I would like to thank an anonymous referee for pointing out that, properly speaking, only detection schemes that use a transfer of energy, such as bar detection schemes, are affected by the hypothesis, and that laser inteferometry (for example) may still be a viable detection technique (see \cite{erp})}.

\section{Conclusions}

In this paper we calculated the energy of a regular black hole of type defined by Mars, Mart\'{\i}n-Prats and Senovilla (a ``MMaS-class'' black hole).  We showed that its energy and momentum could be computed by using the Einstein and Papapetrou energy-momentum complexes, and that these calculations agree with physical intuition.  
Furthermore, we showed that the energy and momentum of a MMaS-class black hole could not be computed in general using the energy-momentum complex due to Landau and Lifshitz, that the energy-momentum complex due to M\o ller gives a less reasonable answer, 
and that these two calculations give results inconsistent with each other and with the energy-momentum complexes of Einstein and Papapetrou.

The results of this paper support the Cooperstock hypothesis.

Chang, Nestor and Chen \cite{CNC1} 
stress the importance of identifying appropriate quasi-local energy-momentum criteria in order to identify 
the right boundary conditions in the definition of energy-momentum density.  
The results of this paper suggest that the Einstein and Papapetrou energy-momentum complexes may define reasonable boundary conditions.  However, the situation is more complicated in general.  For example, the Einstein energy-momentum complex is not symmetric in its indices and so cannot be used to calculate angular momentum.  Furthermore, whereas both the Landau-Lifshitz and the Papapetrou complexes (as defined in this paper) are symmetric in their indices, hence capable of defining conservation laws of angular momentum, the Landau-Lifshitz complex seems to give more reasonable answers in some cases than the Papapetrou complex (see for example \cite{BCJ}).
Additionally, the Papapetrou complex seems much less amenable to calculation for time dependent metrics than its competitors.
In any case the physical meaning of the Einstein boundary term, its apparent dependence on cartesian coordinates, and whether or not the Papapetrou boundary term is deserving of equal footing with Einstein's, is in urgent need of further study.

\section{Acknowledgements}

I thank V. S. Virbhadra for invaluable assistance, and the two anonymous referees for their insightful recommendations.   






\bibliographystyle{plain}

\end{document}